\documentclass[a4paper,twoside]{article}
\usepackage{epsfig}
\usepackage{subcaption}
\usepackage{calc}
\usepackage{amssymb}
\usepackage{amstext}
\usepackage{amsmath}
\usepackage{amsthm}
\usepackage{multicol}
\usepackage{pslatex}
\usepackage{apalike}
\usepackage{SCITEPRESS}     % Please add other packages that you may need BEFORE the 
\usepackage[utf8]{inputenc}
\usepackage{xcolor, soul}
\usepackage{amsmath}
\usepackage{amsthm}
\usepackage{amsfonts}
\usepackage{dsfont}
\usepackage{amssymb}
\usepackage{listings}
\usepackage{url}
\usepackage[ruled, vlined]{algorithm2e}
\usepackage{tikz}
\usepackage{pgfplots}
\pgfplotsset{compat=1.17} 

\usepackage{etoolbox}
\AfterEndPreamble{
  \mathchardef\mathcomma\mathcode`\,
  \mathcode`\,="8000 
}
{\catcode`,=\active
  \gdef,{\mathcomma\discretionary{}{}{}}
}

\begin{document}

\title{Toward Formal Data Set Verification for Building Effective Machine Learning Models}

\author{\authorname{Jorge L\'opez, Maxime Labonne and Claude Poletti}
\affiliation{Airbus Defence and Space, Issy-Les-Moulineaux, France}
\email{\{jorge.lopez-c, maxime.labonne, claude.poletti\}@airbus.com}
}

\keywords{Machine Learning, Data Set Collection, Formal Verification, Trusted Artificial Intelligence}

\abstract{In order to properly train a machine learning model, data must be \emph{properly} collected. To guarantee a proper data collection, verifying that the collected data set holds certain properties is a possible solution. For example, guarantee that data set contains samples across the whole input space, or that the data set is balanced w.r.t. different classes. We present a formal approach for verifying a set of arbitrarily stated properties over a data set. The proposed approach relies on the transformation of the data set into a first order logic formula, which can be later verified w.r.t. the different properties also stated in the same logic. A prototype tool, which uses the z3 solver, has been developed; the prototype can take as an input a set of properties stated in a formal language and formally verify a given data set w.r.t. to the given set of properties. Preliminary experimental results show the feasibility and performance of the proposed approach, and furthermore the flexibility for expressing properties of interest.}

\onecolumn \maketitle \normalsize \setcounter{footnote}{0} \vfill

\maketitle

\section{\uppercase{Introduction}}\label{sec:intro}
In the past few decades, Machine Learning~(ML) has gained a lot of attention, partially due to the creation of software libraries (e.g., \cite{scikit-learn}) that ease the usage of complex algorithms. In this context, the volume of stored data has dramatically increased over the last few years. However, an often overlooked task is the data extraction and collection to create proper data sets to train efficient machine learning models. 

When retrieving information for the data set collection, there are key points to take into consideration. The reason is that ML models generalize their output based on the training (seen) data. However, a problem that is commonly encountered is that a model is expected to generalize well unseen regions of the input space while such regions do not behave in accordance to the provided training data. Another problem that often occurs is that there is a class in the data set which is underrepresented (e.g., for an anomaly detection data set, 99\% of the examples are normal events). In general, many data biases can occur in a collected data set. A simple strategy while collecting data sets is to collect a large number of entries, conjecturing that important data are likely to be found if more data are available. However, this strategy yields incorrect results, and moreover, large data sets can cause ML models to be trained for longer than necessary; this in turn can make certain algorithms which may yield accurate results unusable for such cases. Additionally, with the proliferation of machine generated data sets, for example via Generative Adversarial Networks, assuring that the generated data set holds some properties of interest is of utmost importance.

In order to guide the collection of a proper data set to effectively train a ML model, verifying that a partially collected data set holds certain properties of interest is a possible solution. This verification can be done with the use for formal methods, such as deductive verification \cite{smt1}. Considering a formal specification of a data set, a formal proof that the data set holds certain properties can be provided. Whenever this specification is violated (certain properties do not hold), identifying the properties that do not hold may help to diagnose the missing or incorrect information. This paper is devoted to the formal verification of machine learning data sets through the use of Satisfiability Modulo Theories~(SMT) \cite{smt1} (for preliminary concepts on ML and SMT, see Section~\ref{sec:prelim}). The approach is based on the encoding of a data set into a Many-Sorted First Order Logic~(MSFOL) formula which is later verified together with the desired set of properties (see Section~\ref{sec:verif}). 

A tool for the verification of data sets has been developed. The tool relies on the use of the widely-known z3 \cite{z3} solver. Preliminary experimental results show that in spite of the high computational complexity of SMT procedures, for the verification of data sets, these properties can be verified in a reasonable amount of time (see Section~\ref{sec:experim}). 

It is important to note that verifying certain properties over a data set is a task which is consistently considered as necessary, and a norm for many practitioners. However, in the literature very few researchers focus on automatic validation of data sets (see for example \cite{dsvalid}). Furthermore, to the best of our knowledge, there is no work which aims at providing means for the verification of arbitrarily stated properties, and moreover, in a formal manner. In this light, this paper aims at exploring this direction.

\section{\uppercase{Preliminaries}}\label{sec:prelim}
\input{smt-highlight}
In order to make our paper as self-contained as possible, we have included a brief description of some preliminary concepts required in our work. 

\subsection{Machine learning and structured data sets}
We consider that a \emph{structured machine learning data set} contains \emph{examples} alongside with their \emph{expected outputs}. Given the inputs and expected outputs, the final goal of a supervised ML algorithm is to learn how to map a training example to its expected output. For an unsupervised ML algorithm the goal is to learn patterns from the data; thus, the expected output does not exist. In our work, we consider that the expected outputs are always present, and thus, a data set for unsupervised machine learning (where there are no expected outputs) has the same expected output for all training examples. Further, we consider only structured data sets.

Formally, the inputs are called \emph{features} or \emph{parameters}. A \emph{feature vector}, denoted as $\mathbf{X}$, is an $n$-tuple of the different inputs, $x_{1}, x_{2}, \dots, x_{n}$. The expected output for a given feature vector is called a \emph{label}, denoted simply as $y$, and the possible set of outputs is respectively denoted as $Y$. The set of examples, called a \emph{training data set}, consists of pairs of a feature vector and a label; each pair is called a \emph{training example}, denoted as $(\mathbf{X}, y)$. For convenience, we represent the data set as a matrix $D_{m\times n}$ and a vector $O_{m}$ where $D$ contains the feature vectors and $O$ contains the expected outputs for a data set of cardinality $m$. The vector representing the $i$-th row (training vector) is denoted as $D_{i}$, and its associated expected output as $O_i$. Likewise, the $j$-th feature (column vector) is denoted as $D^{T}_{j}$ ($D^T$ denotes the transpose of the matrix $D$). Finally, the $j$-th parameter of the $i$-th training example is denoted by the matrix element $d_{i,j}$.

\subsection{Satisfiability Modulo Theories~(SMT)}
SMT is a decision problem, that for a given first order logic formula $\phi$ searches if $\phi$ is satisfiable w.r.t. a set of background theories. For example, w.r.t. integer linear arithmetic, the following formula is satisfiable: $\Phi =  (x \in \mathbb{Z}) \wedge  (y \in \mathbb{Z}) \wedge (x < y) \wedge (x < 0) \wedge (y > 0) \wedge (x + y > 0)$; the formula can be satisfied for instance by the interpretation $x=-1,y=2$. The importance of restricting an interpretation of certain function and predicate symbols in a first-order logic formula (according to a background theory $\mathcal{T}$), is that specialized decision procedures have been proposed; thus, making the problem of checking the satisfiability of such formulas decidable.

It is important to note that many of the applications that use SMT involve different data types \cite{smt1}. Therefore, SMT usually works with a \emph{sorted} (typed) version of first order logic \cite{sortedlogic}. Essentially, in SMT there exists a finite set of sort symbols (types) $S$ and an infinite set of variables $X$ for the (sorted) formulas, where each variable has a unique associated sort in $S$. This is an oversimplification of a many-sorted first order logic~(MSFOL). As MSFOL is useful to express our formulas of interest, in the next subsection we provide a formal definition of its syntax \cite{msfol,smt1,smt2}. 

\subsubsection{Many-sorted First-order Logic Syntax}

A \emph{signature} is a tuple $\Sigma=(S,C,F,P)$, where $S$ is a non-empty and finite set of sorts, $C$ is a countable set of constant symbols whose sorts belong to $S$, $F$ and $P$ are countable sets of function and predicate symbols correspondingly whose arities are constructed using sorts that belong to $S$. Predicates and functions have an associated arity in the form $\sigma_1\times\sigma_2\times\ldots\times\sigma_n\rightarrow\sigma$, where $n\geq1$ and $\sigma_1,\sigma_2,\ldots,\sigma_n,\sigma \in S$. 

A \emph{$\Sigma$-term} of sort $\sigma$ is either: (i) each variable $x$ of sort (type) $\sigma$, where $\sigma\in S$; (ii) each constant $c$ of sort (type) $\sigma$, where $\sigma\in S$; and (iii) $f\in F$ with arity $\sigma_1\times\sigma_2\times\ldots\times\sigma_n\rightarrow\sigma$, is a term of sort $\sigma$, thus, for $f(t_1,\ldots,t_n)$, $t_i$ (for $i \in \{1,\ldots,n\}$) is a $\Sigma$-term of sort $\sigma_i$. 

A \emph{$\Sigma$-atom} ($\Sigma$-atomic formula) is an expression in the form $s=t$ or $p(t_1,t_2,\ldots,t_n)$, where $=$ denotes the equality symbol, $s$ and $t$ are $\Sigma$-terms of the same sort, $t_1,t_2,\ldots,t_n$ are $\Sigma$-terms of sort $\sigma_1,\sigma_2,\ldots,\sigma_n \in S$, respectively, and $p$ is a predicate of arity $\sigma_1\times\sigma_2\times\ldots\times\sigma_n$.

A \emph{$\Sigma$-formula} is either: (i) a $\Sigma$-atom; (ii) if $\phi$ is a $\Sigma$-formula, $\neg\phi$ is a $\Sigma$-formula, where $\neg$ denotes negation; (iii) if both $\phi, \psi$ are $\Sigma$-formulas, then, $\phi\wedge\psi$ and $\phi\vee\psi$ are $\Sigma$-formulas (likewise, the short notations $\phi\rightarrow\psi$ and $\phi\leftrightarrow\psi$ for $\neg\phi\vee\psi$ and $(\phi\wedge\psi) \vee (\neg\phi\wedge\neg\psi)$); finally, (iv) if  $\phi$ is a $\Sigma$-formula and $x$ is a variable of sort $\sigma$, then, $\exists x\in\sigma\; \phi$ ($x\in\sigma$ is used to indicate that $x$ has the sort $\sigma$) is a $\Sigma$-formula (likewise, the short notation $\forall x\in\sigma\; \phi$ for $\neg\exists x\in\sigma\; \neg\phi$), where $\exists$ denotes the existential quantifier and $\forall$ denotes the universal quantifier, as usual.

 We leave out the formal semantics of MSFOL formulas, their interpretations and satisfiability as we feel it can unnecessarily load the paper with unused formalism. However, we briefly discuss some aspects of MSFOL formula satisfiability. As previously mentioned, for some signatures, there exist decision procedures, which help to determine if a given formula is satisfiable. For example, consider the signature with a single sort $\mathds{R}$, all rational number constants, functions ${+,-,*}$ and the predicate symbol $\leq$; SMT will interpret the constants, symbols and predicates as in the usual real arithmetic sense $\mathds{R}$. The satisfiability of $\Sigma$-formulas for this theory (real arithmetic) is decidable, even for formulas with quantifiers \cite{smt1,decision_proc_survey}, i.e., for some infinite domain theories, there exist procedures\footnote{Often such procedures seek to ``eliminate'' the quantifiers and obtain an equivalent quantifier-free formula} to decide if a given quantified formula is satisfiable. Therefore, the satisfiability for formulas as: $\exists n\in\mathds{R}\; \forall x\in\mathds{R}\; x+n=x $ can be automatically determined (via a computer program implementing the decision procedure, i.e., an SMT solver). If a formula is satisfiable, there exists an interpretation (or model) for the formula, i.e., a set of concrete values for the variables, predicates and functions of the formula that makes this formula evaluate to \textsc{true}.

\section{\uppercase{Data set encoding and formal verification}}\label{sec:verif}
As previously mentioned (see Section~\ref{sec:prelim}), a ML data set is composed of a matrix $D_{m\times n}$ and a vector $O_m$, where $m$ is the number of training examples, $n$ the number of features, $D$ contains the training examples, and $O$ the expected outputs. However, note that in our definition of this matrix we never mentioned the type of each feature in the data set. In general, there is no theoretical limitation over the type of these features, nonetheless, for practical reasons, we consider that all features are real valued. The main reason is that otherwise additional information would be required for each of the features. Moreover, in practice, well-known libraries work with real-valued features. As usual, for those features which are not naturally real, an encoding must be found (for example, one hot encoding for categorical features, etc.). Thus, we consider that $d_{i,j},o_i\in \mathds{R}\;\forall i\in\{1,\ldots,m\}, j\in\{1,\ldots,n\}$. Additionally, we assume that $O$ is always present in the data sets, independently if this data set is meant for supervised or unsupervised machine learning. If a data set is not \emph{labeled}, then $\forall i,k\in\{1,\ldots,m\}\;o_i=o_k$.

\paragraph{Encoding a ML dataset as a MSFOL formula.} Having a convenient formal description for a data set eases the encoding of this data set as a MSFOL formula. To encode the data as a formula, we make use of the theory of arrays\footnote{The theory of arrays considers basic read and write axioms.}. We denote that an object $a$ is of sort array with indices of type (sort) $\mathcal{T}1$ and holding objects of type $\mathcal{T}2$ as $a\in \mathds{A}_{\mathcal{T}1,\mathcal{T}2}$. Indeed, a data set can be encoded using Algorithm~\ref{algo:ds_enc}.

\begin{algorithm}[!htb]
%\small
\SetKwInOut{Input}{Input}\SetKwInOut{Output}{Output}\SetKw{KwBy}{b}\SetKw{Goto}{go to}

    \Input{A data set $D_{M\times N}$ (with $N$ features and $M$ training examples), and its expected output vector $O_M$}
    \Output{A MSFOL formula representation of the data set $\phi$}
    \textbf{Step 0:} Set $\phi\leftarrow\text{\textsc{true}}$, set $labels\leftarrow$\textsc{Array}(), and set $L\leftarrow0$\;
    \textbf{Step 1:} Set $\phi\leftarrow\phi\wedge(m,n,l\in\mathds{Z})\wedge(m=M)\wedge(n=N)$\;
    \textbf{Step 2:} Set $\phi\leftarrow\phi\wedge(\mathcal{D}\in \mathds{A}_{\mathds{Z},\mathds{A}_{\mathds{Z},\mathds{R}}})\wedge(\mathcal{O}\in\mathds{A}_{\mathds{Z},\mathds{R}})\wedge(\mathcal{L}\in\mathds{A}_{\mathds{Z},\mathds{R}})$\;
    \textbf{Step 3:} \For{$i\leftarrow0; i < M; i\leftarrow i + 1$}
    {
        Set $add\leftarrow$\textsc{true}\;
        \For{$j\leftarrow0; i < N; j\leftarrow j + 1$}
        {
            Set $\phi\leftarrow\phi\wedge(\mathcal{D}[i][j] = d_{i,j})$\;
        }
        Set $\phi\leftarrow\phi\wedge(\mathcal{O}[i] = o_{i})$\;
        \For{$k\leftarrow 0; k < L; k\leftarrow k + 1$}
        {
            \If{$labels[k] = o_i$}
            {
                Set $add\leftarrow$\textsc{false}\;
            }
        }
        \If{$add$}
        {
            Set $labels[L]\leftarrow o_i$\; 
            Set $\phi\leftarrow\phi\wedge(\mathcal{L}[L] = o_{i})$\;
            Set $L\leftarrow L + 1$\;
        }
    }
    \textbf{Step 4:} Set $\phi\leftarrow\phi\wedge(l=L)$ and \Return{$\phi$}
    \caption{Data set encoding}\label{algo:ds_enc}
\end{algorithm}

%formal verification of data sets properties and data set specification definiton, satisfiability of the specification
\subsection{Formal verification of data sets}
Indeed, a data set can be formally defined as an MSFOL formula $\phi_{ds}$ which holds the following properties: $\phi_{ds}$ is a conjunction of \emph{five} main parts, that is, i) the assertion that an integer variable $m$ is of the size of the number of training examples, a variable $n$ is of the size of the features and a variable $l$ is of the size of the distinct labels, ii) the assertion that $\mathcal{D}$ is a two-dimensional (integer indexed) real-valued array (of size $m\times n$) and $\mathcal{O}, \mathcal{L}$ are integer indexed real-valued arrays (of size $m$, and $l$, respectively) iii) $\mathcal{D}[i][j]$ contains the $j$-th feature value for the $i$-th training example; iv) $\mathcal{O}[i]$ contains the expected output for the $i$-th training example; and, v) $\mathcal{L}[i]$ contains the $i$-th (distinct) label.

We assume that we want to verify $k$ properties over the data set, and furthermore, that these properties are expressed also in MSFOL. Indeed, MSFOL allows to express many properties of interest (in Section~\ref{sec:ex_prop} we showcase its expressiveness). Therefore, we assume that we are given $\pi_1,\ldots,\pi_k$ MSFOL formulas to verify. These properties involve the variables in $\phi_{ds}$. Additionally, we assume that these formulas should all \emph{hold} independently over the data set, and their conjunction is \emph{satisfiable}. Thus, impose a restriction that $\pi_x\wedge\pi_y$ is satisfiable, for $x,y\in\{1,\ldots,k\}$; we call this set of properties the \emph{data set specification} $\sigma$. This means that two properties may not \emph{contradict} each other. For example, it cannot be required that the data set has more than 30 training examples and at the same time that it must have at most 20 ($(\pi_1\leftrightarrow (m> 30))\wedge(\pi_2\leftrightarrow (m\leq 20))$). Additionally, the conjunction of properties must be satisfiable means that there is an interpretation that makes this formula (the conjunction) evaluate to \textsc{true}, i.e., there exists a data set which can satisfy this specification. Otherwise, the verification of any data set is useless as no data set can hold such set of properties.

\paragraph{The formal data set verification problem} can be reduced to the following: given a data set formula $\phi_{ds}$ (created using Algorithm~\ref{algo:ds_enc} from $D$ and $O$) and a data set specification $\sigma=\bigwedge_{l=1}^{k}\pi_l$, is $\phi_{ds}\wedge\sigma$ satisfiable? If the conjunction of these formulas is satisfiable then, each of the properties must hold for the data set as the conjunction of all properties is satisfiable by itself; if the conjunction is satisfiable we say that the data set \emph{holds} the properties $\pi_1,\ldots,\pi_k$ or that the data set \emph{conforms} to the specification $\sigma$. Perhaps this is quite an abstract view of the problem. For that reason, in the following subsection we provide concrete examples that should help the reader to better understand.

\subsection{Example data set and properties}\label{sec:ex_prop}
First, let us consider a very small data set:
\[D=\begin{pmatrix}0.051267 & 0.69956\\
-0.092742 & 0.68494\\   
-0.21371 & 0.69225\\
-0.375 & 0.50219\\
-0.51325 & 0.46564\\
-0.52477 & 0.2098\\
-0.39804 & 0.034357\\
-0.30588 & -0.19225\\
0.016705 & -0.40424\\
0.13191 & -0.51389\end{pmatrix},
O=\begin{pmatrix} 1\\ 
 0\\ 
 -1\\ 
 -1\\ 
 -1\\ 
 -1\\ 
 -1\\ 
 -1\\ 
 -1\\ 
 -1\\ \end{pmatrix}\]

After applying Algorithm~\ref{algo:ds_enc} to $D$ and $O$ as shown before, the output ($\phi_{ds}$) is: \[\begin{aligned}
&(m,n,l\in\mathds{Z})\wedge(m=10)\wedge(n=2)\wedge\\
&(\mathcal{D}\in \mathds{A}_{\mathds{Z},\mathds{A}_{\mathds{Z},\mathds{R}}})\wedge
(\mathcal{O}\in\mathds{A}_{\mathds{Z},\mathds{R}})\wedge(\mathcal{L}\in\mathds{A}_{\mathds{Z},\mathds{R}})\\
&\wedge(\mathcal{D}[0][0] = 0.051267)\wedge(\mathcal{D}[0][1] = 0.69956)\\
&\wedge(O[0]=1)\wedge(\mathcal{L}[0]=1)\\
&\wedge(\mathcal{D}[1][0] = -0.092742)\wedge(\mathcal{D}[1][1] = 0.68494)\\
&\wedge(O[1]=0)\wedge(\mathcal{L}[1]=0)\\
&\wedge(\mathcal{D}[2][0] = -0.21371)\wedge(\mathcal{D}[2][1] = 0.69225)\\
&\wedge(O[2]=-1)\wedge(\mathcal{L}[2]=-1)\\
&\wedge(\mathcal{D}[3[0] = -0.375)\wedge(\mathcal{D}[3][1] = 0.50219)\\
&\wedge(O[3]=-1)\\
&\wedge(\mathcal{D}[4][0] = -0.51325)\wedge(\mathcal{D}[4][1] = 0.46564)\\
&\wedge(O[4]=-1)\\
&\wedge(\mathcal{D}[5][0] = -0.52477)\wedge(\mathcal{D}[5][1] = 0.2098)\\
&\wedge(O[5]=-1)\\
&\wedge(\mathcal{D}[6][0] = -0.39804)\wedge(\mathcal{D}[6][1] = 0.034357)\\
&\wedge(O[6]=-1)\\
&\wedge(\mathcal{D}[7][0] = -0.30588)\wedge(\mathcal{D}[7][1] = -0.19225)\\
&\wedge(O[7]=-1)\\
&\wedge(\mathcal{D}[8][0] = 0.016705)\wedge(\mathcal{D}[8][1] = -0.40424)\\
&\wedge(O[8]=-1)\\
&\wedge(\mathcal{D}[9][0] = 0.13191)\wedge(\mathcal{D}[9][1] = -0.51389)\\
&\wedge(O[9]=-1)\wedge(l=3)\end{aligned}\]

Let us start by showcasing very simple properties and how their formal verification works. Suppose the specification consists of a single property: ``the data set must contain at least 100 training examples,'' this property can be expressed in MSFOL simply as $\pi_\#\leftrightarrow(m\geq100)$. Notice how $\phi_{ds}\wedge\pi_\#$ is not satisfiable as there does not exist an interpretation that makes it evaluate to \textsc{true}; particularly, if $m$ is greater than 99, then the clause (in $\phi_{ds}$) $m=10$ cannot evaluate to \textsc{true} and since this is a conjunction, $\phi_{ds}\wedge\pi_\#$ evaluates to \textsc{false}. Similarly, if $m$ is 10, then the $\pi_\#$ makes the conjunction evaluate to \textsc{false}. Thus, we say that the data set does not hold the property $\pi_\#$.

Let us start examining more complex properties that can be formally verified over the data set. A slightly more complex property to verify is: ``the data set must be min-max normalized,'' which can be expressed in MSFOL as  $\pi_\pm\leftrightarrow\nexists (i,j\in\mathds{Z}) ( (i \geq 0) \wedge (i < n) \wedge (j \geq 0) \wedge (j < m) \wedge ((\mathcal{D}[i][j] < min)\vee (\mathcal{D}[i][j] > max)) )$. Certainly $min$ and $max$ are defined constants (e.g., -1 and 1) an either these variables must be defined or the value must be replaced; for $min=-1$ and $max=1$, $\phi_{ds}$ holds the property $\pi_\pm$ (as $\phi_{ds}\wedge\pi_\pm$ is satisfiable).

The previous properties are useful to showcase how easy is to translate desired properties into the formalism. However, verifying such properties is quite simple, and furthermore can be uninteresting as the data set can be normalized later on, for example. As previously stated, our motivation comes from the proper extraction and collection of the data set. We have discussed the case where training examples are provided for some regions of the input space and some other regions are overlooked. To verify that ``the data set is sampled across the whole input space,'' the following property can be verified $\pi_\ast\leftrightarrow\nexists (p\in\mathcal{A}_{\mathds{Z},\mathds{R}}) \forall (i\in\mathds{Z}) ((i \geq 0) \wedge (i< m)) \implies (\sqrt{\sum_{j=0}^{m-1}(p[j]-\mathcal{D}[i][j])^2}>\delta)$; the property basically states that there does not exist a point such that it has a greater Eucledian distance that a chosen constant $\delta$. As an example, for $\delta=1$, our example data set does not hold the previous property $\pi_\ast$ as there exists a point in the input space that has greater Eucliden distance, for example if $p[0]=2$ and $p[1]=2$. Note that the property never specifies the minimum or maximum values of the input space and thus, it is likely that no data set is sampled over an infinite domain. An easy solution is to add such constraints to $\pi_\ast$, i.e., $\nexists (l\in\mathds{Z}\wedge(l\geq 0)\wedge(l<n)(\wedge(p[l]>max)\vee(p[l]<min)))$, for given $max$ and $min$ constants. We draw the reader's attention to the fact that a formal specification must be well-stated and this is an assumption of our work and generally in any formal verification strategy.

Finally, note that sometimes it is more convenient to state negated properties. For example, to verify that the data set is \emph{balanced}, we can verify the following property: ``there is no class which has less than $\frac{m}{(\beta*l)}$ samples,'' where $l$ is the number of different outputs (labels) and $\beta$ is a chosen constant. This property states that the data set must have equal amount of samples, up to a given constant. For example, if $\beta=1$ the data set must be perfectly balanced, while if $\beta=2$ only half of the samples (of a perfectly balanced data set) are required per class. It is important to state that unbalanced data sets represent a real problem for current machine learning algorithms, and moreover, it is often encountered in the domain. Accordingly, researchers actively try to tackle the problem (see for example \cite{imb}). Indeed, it can be not that intuitive how to state this property in first order logic. There are many particularities that must be considered; for example, the fact that there is no notion of loops in first order logic and we require to define a function to count the number of instances where a given label appears. To overcome this particular problem a recursive function can be defined. In order to keep the paper readable, we avoid this definitions and simply denote defined functions in mathematical bold-font. The interested reader can refer to the prototype implementation section (Section~\ref{sec:experim}) and correspondingly to the tool's repository to check the full property implementation. We state the aforementioned property as: $\pi_\equiv\leftrightarrow\nexists i\in\mathds{Z} ((i \geq 0)\wedge (i<l)\wedge(\mathbf{S}(\mathcal{O},\mathcal{L}[i], m) < \frac{m}{\beta*l} ))$, where $\mathbf{S}(A, v, s)$ is a function that returns the number of times the value $v$ is found in an array $A$ up to index $s$; that is, that is how many times the label is found in the label array.

We have exemplified different properties that can be formally verified in data sets. We do not focus on an extensive list of properties but, rather on providing means for formally verifying any property in a given data set. We could state much more properties, for example, there are no \emph{contradicting training examples} in the data set, i.e., there does not exist two equal elements in $D$ with different indices for which the corresponding elements in $O$ differ. We limit this section with these examples. However, we note that as shown in the previous examples, the formalism is quite flexible for expressing real properties of interest.

\section{\uppercase{Tool development and experimental results}}\label{sec:experim}
In order to assess the feasibility and efficiency of the proposed approach, a prototype tool has been developed in Julia \cite{julia}. Generally, speaking, the tool takes as an input: a Comma Separated Values~(CSV) file as a data set, assuming that the last column of each row must be the expected output for the training example (remainder of the columns); a directory, where the properties to be checked are stored, one per file in the \verb|SMT-LIB| language.

\paragraph{SMT-LIB} is a language that many SMT solvers can take as an input and its syntax is quite intuitive. For example, for expressing the property $\nexists (i,j\in\mathds{Z}) ( (i \geq 0) \wedge (i < n) \wedge (j \geq 0) \wedge (j < m) \wedge ((\mathcal{D}[i][j] < min)\vee (\mathcal{D}[i][j] > max)) )$ can be simply done in SMT-LIB as shown in Listing~\ref{lst:smtlib}. 

\begin{lstlisting}[language=smtlib, label=lst:smtlib, caption={$\pi_\pm$ in SMT-LIB}]
(assert
 (not
  (exists ((i Int) (j Int))  
   (and
    (>= i 0)
    (< i n)
    (>= j 0)
    (< j m)
    (or
     (< (select (select D i) j) min )
     (> (select (select D i) j) max )
    )
   )
  )
 )
)
\end{lstlisting}

The tool works as described in Algorithm~\ref{algo:tool}. Note that, $SMT$ is an SMT procedure call to determine if the given formula is satisfiable. In our tool, we use the z3 \cite{z3} solver (which takes as an input the SMT-LIB format). The interested reader can check the properties stated in SMT and more information about our tool in the tool's repository \cite{repo}.

\begin{algorithm}[!htb]
%\small
\SetKwInOut{Input}{Input}\SetKwInOut{Output}{Output}\SetKw{KwBy}{b}\SetKw{Goto}{go to}

    \Input{A CSV data set file $f$ (with $n\geq1$ features, and $m\geq1$ training examples), and a directory $d$ containing property files}
    \Output{Verdicts for each property $\pi\in d$}
    \textbf{Step 0:} Read $f$ and store it into the arrays $D$ and $O$, and set $m$ and $n$, correspondingly\;
    \textbf{Step 1:} Use Algorithm~\ref{algo:ds_enc} to obtain $\phi_{ds}$ from $D,O,m,$ and $n$\;
    \textbf{Step 2:} \ForEach{$p\in d$ }
    {
        Read the contents of $p$ into the formula $\pi$\;
        \If{$SMT(\phi_{ds}\wedge\pi)$ is satisfiable}
        {
            $display(\pi$ holds in the data set $f)$ 
        }
        \Else
        {
            $display(\pi$ does not hold for the data set $f)$
        }
    }
    \caption{Data Set Verification}\label{algo:tool}
\end{algorithm}

\subsection{Preliminary experimental results}
All experiments were executed with commodity hardware with the intention to showcase the performance of the proposed approach. The experiments were performed with an Ubuntu 20.04LTS with 4 Intel(R) Core(TM) i5-6300U CPU @ 2.40GHz, and 8GB of RAM. 

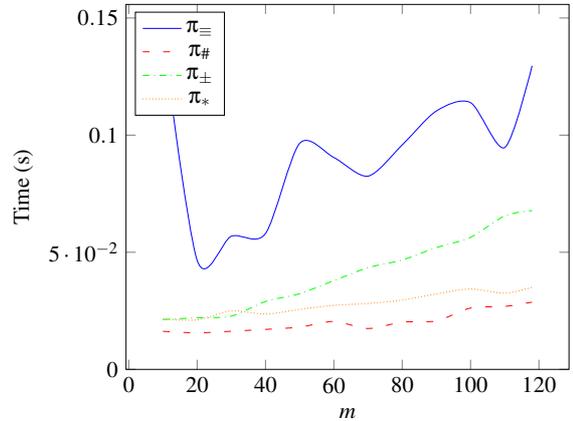
\begin{figure}[!htb]
    \centering
    \begin{tikzpicture}[scale=.85]
\begin{axis}[
    ymin=0,
	ylabel={Time (s)},
	xlabel={$m$},
	legend style={at={(0.02,0.98)}, 
	anchor=north west},
]

\addplot[smooth, solid, blue] coordinates
{
    (10, 0.141627)
    (20, 0.046552)
    (30, 0.056845)
    (40, 0.058008)
    (50, 0.096480)
    (60, 0.090452)
    (70, 0.082492)
    (80, 0.095818)
    (90, 0.110307)
    (100, 0.113863)
    (110, 0.094700)
    (118, 0.129609)
};\addlegendentry{$\pi_\equiv$}

\addplot[smooth, loosely dashed, red] coordinates
{
    (10, 0.016247)
    (20, 0.015563)
    (30, 0.016290)
    (40, 0.017066)
    (50, 0.018053)
    (60, 0.020448)
    (70, 0.017457)
    (80, 0.020176)
    (90, 0.020702)
    (100, 0.026182)
    (110, 0.026866)
    (118, 0.028705)
};\addlegendentry{$\pi_\#$}

\addplot[smooth, dashdotted, green] coordinates
{
    (10, 0.021307)
    (20, 0.022026)
    (30, 0.022748)
    (40, 0.028983)
    (50, 0.032315)
    (60, 0.037894)
    (70, 0.043429)
    (80, 0.046728)
    (90, 0.051970)
    (100, 0.056375)
    (110, 0.065533)
    (118, 0.067794)
};\addlegendentry{$\pi_\pm$}

\addplot[smooth, densely dotted, orange] coordinates
{
    (10, 0.021453)
    (20, 0.021112)
    (30, 0.024985)
    (40, 0.023605)
    (50, 0.025602)
    (60, 0.027347)
    (70, 0.028128)
    (80, 0.029578)
    (90, 0.032225)
    (100, 0.034313)
    (110, 0.032531)
    (118, 0.035049)
};\addlegendentry{$\pi_\ast$}

\end{axis}
\end{tikzpicture}
    \caption{Performance of formal data set verification}
    \label{fig:perf}
\end{figure}

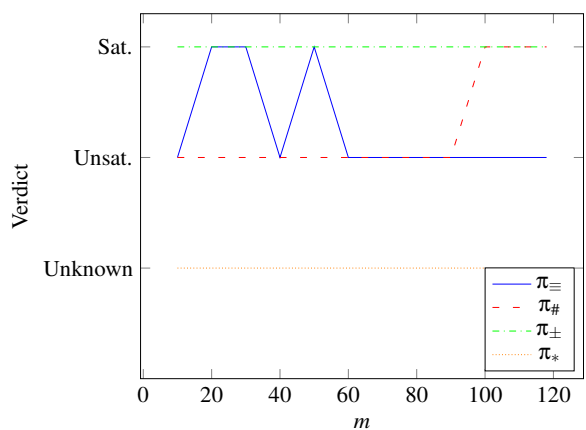
\begin{figure}[!htb]
    \centering
    \begin{tikzpicture}[scale=.85]
\begin{axis}[
    ymin=0,
    ytick={1,2,3},
    yticklabels = {Unknown, Unsat., Sat.},
	ylabel={Verdict},
	xlabel={$m$},
	legend style={at={(0.98,0.02)}, 
	anchor=south east},
]

\addplot[solid, blue] coordinates
{
    (10, 2)
    (20, 3)
    (30, 3)
    (40, 2)
    (50, 3)
    (60, 2)
    (70, 2)
    (80, 2)
    (90, 2)
    (100, 2)
    (110, 2)
    (118, 2)
};\addlegendentry{$\pi_\equiv$}

\addplot[ loosely dashed, red] coordinates
{
    (10, 2)
    (20, 2)
    (30, 2)
    (40, 2)
    (50, 2)
    (60, 2)
    (70, 2)
    (80, 2)
    (90, 2)
    (100, 3)
    (110, 3)
    (118, 3)
};\addlegendentry{$\pi_\#$}

\addplot[dashdotted, green] coordinates
{
    (10, 3)
    (20, 3)
    (30, 3)
    (40, 3)
    (50, 3)
    (60, 3)
    (70, 3)
    (80, 3)
    (90, 3)
    (100, 3)
    (110, 3)
    (118, 3)
};\addlegendentry{$\pi_\pm$}

\addplot[densely dotted, orange] coordinates
{
    (10, 1)
    (20, 1)
    (30, 1)
    (40, 1)
    (50, 1)
    (60, 1)
    (70, 1)
    (80, 1)
    (90, 1)
    (100, 1)
    (110, 1)
    (118, 1)
};\addlegendentry{$\pi_\ast$}

\end{axis}
\end{tikzpicture}
    \caption{Satisifiability of properties (w.r.t the data set conjunction)}
    \label{fig:sat}
\end{figure}

In order to evaluate the feasibility of our proposed solution, the properties $\pi_\#,\pi_\pm, \pi_\ast$ and $\pi_\equiv$ have been encoded in SMT-LIB, and a data set was incrementally tested. We present the results of both the performance and satisfiability of properties w.r.t. the data sets in Figures~\ref{fig:perf}, \ref{fig:sat}, respectively. As can be seen, the performance of the proposed approach is acceptable; as any formal verification approach, the decision procedures are often exponential in the worst case. For formally guaranteeing that the data set holds certain properties of interest, this procedure can be executed once, in which case the running time is not much of a constraint. Our preliminary experimental evaluation shows that properties are solved fast (milliseconds per hundreds of training examples), specially simple properties (e.g., $\pi_\#$).

It is interesting to observe the satisfiability of the properties. It is normal that when adding more training examples the data set may get balanced or unbalanced ($\pi_\equiv$); it is also normal that all data sets which have less than 100 training examples fail the property $\pi_\#$. One can conclude that the example data set is also well min/max normalized as $\pi_\pm$ is always satisfiable. Finally, note that even if the language allows it and solver can read the property $\pi_\ast$, the property is very complicated as it is quantified over an array; the solver cannot process such complex formulation and so the property always returns an unknown status. We envision different strategies to overcome this problem. For example, instead of formulating the problem as it is, to pre-process the dimension of the training vector, and ask the formula quantified over $n$ reals ($\exists p_1,\ldots,p_n\in\mathds{R}\psi$). This should effectively reduce the complexity of the formula, however, this may require a Domain Specific Language~(DSL) for stating properties of interest. However, note that this approach is out of the scope of this initial study.

\section{\uppercase{Conclusion and future work}}\label{sec:conc}
%future work, automatically repairing the data sets

In this paper, we have proposed a formal data set verification approach. Such formal verification can be used for guaranteeing that the data extraction is adequate for properly training a machine learning model. We have showcased different formal properties to be verified over the data sets, and experimentally proven that the approach is feasible, and furthermore flexible w.r.t. the semantic capabilities of the proposed formalism. 

As for future work we plan to test the performance of the approach on large scale data sets. Also, we intend to further investigate DSLs for property specification (as discussed in Section~\ref{sec:experim}). Additionally, as each of the training examples gets translated into a part of a formula, it is interesting to try to remove some training examples when a property is not satisfiable in order to obtain a satisfiable one; this would allow to automatically repair data sets w.r.t. a set of properties. Nevertheless, different elements must be taken into consideration, for example, the fact that the model found by the solver may include other training examples, which are \emph{fictitious}. Finally, an interesting direction is to consider the formal verification of unstructured data for machine learning.

\bibliographystyle{apalike}
{\small
\bibliography{references}}

%\clearpage
%\section*{Reviewers' comments}
%\input{reviews}

\end{document}